\documentclass[aps,prx,longbibliography,notitlepage,reprint,superscriptaddress]{revtex4-2}

\usepackage[colorlinks,linkcolor=blue,citecolor=blue,urlcolor=red]{hyperref}

\usepackage{soul}
\usepackage{changes}

\usepackage{float}
\usepackage{CJK}
\usepackage{graphicx}
\usepackage{amsmath,amssymb,amsfonts}
\usepackage{siunitx}
\usepackage{soul}

\usepackage{amsmath,amssymb,amsfonts} 
\usepackage{bm}	
\renewcommand{\mathbf}{\bm}

\usepackage{dsfont}	
\renewcommand{\mathbb}{\mathds}	

\usepackage{mathrsfs} 
\usepackage{mathtools} 

\newcommand{\fref}[2]{Fig.~\hyperref[#1]{\ref{#1}#2}}
\renewcommand{\eqref}[1]{Eq.~(\ref{#1})}

\newcommand{\SiN}{$\mathrm{Si_3N_4}$}

\begin{document}

\author{Mohammad. J. Bereyhi}
\affiliation{Institute of Physics (IPHYS), Swiss Federal Institute of Technology Lausanne (EPFL), 1015 Lausanne, Switzerland}

\author{Amirali Arabmoheghi}
\affiliation{Institute of Physics (IPHYS), Swiss Federal Institute of Technology Lausanne (EPFL), 1015 Lausanne, Switzerland}

\author{Sergey A. Fedorov}
\affiliation{Institute of Physics (IPHYS), Swiss Federal Institute of Technology Lausanne (EPFL), 1015 Lausanne, Switzerland}

\author{\\Alberto Beccari}
\affiliation{Institute of Physics (IPHYS), Swiss Federal Institute of Technology Lausanne (EPFL), 1015 Lausanne, Switzerland}

\author{Guanhao Huang}
\affiliation{Institute of Physics (IPHYS), Swiss Federal Institute of Technology Lausanne (EPFL), 1015 Lausanne, Switzerland}

\author{Tobias J. Kippenberg}
\email{tobias.kippenberg@epfl.ch}
\affiliation{Institute of Physics (IPHYS), Swiss Federal Institute of Technology Lausanne (EPFL), 1015 Lausanne, Switzerland}

\author{Nils J. Engelsen}
\email{nils.engelsen@epfl.ch}
\affiliation{Institute of Physics (IPHYS), Swiss Federal Institute of Technology Lausanne (EPFL), 1015 Lausanne, Switzerland}

\title{Nanomechanical resonators with ultra-high-$Q$ perimeter modes}

	
\begin{abstract}

Systems with low mechanical dissipation are extensively used in precision measurements such as gravitational wave detection, atomic force microscopy and quantum control of mechanical oscillators via opto- and electromechanics. The mechanical quality factor ($Q$) of these systems determines the thermomechanical force noise and the thermal decoherence rate of mechanical quantum states. While the dissipation rate is typically set by the bulk acoustic properties of the material, by exploiting dissipation dilution, mechanical $Q$ can be engineered through geometry and increased by many orders of magnitude. Recently, soft clamping in combination with strain engineering has enabled room temperature quality factors approaching one billion ($10^9$) in millimeter-scale resonators. Here we demonstrate a new approach to soft clamping which exploits vibrations in the perimeter of polygon-shaped resonators tethered at their vertices. In contrast to previous approaches, which rely on cascaded elements to achieve soft clamping, perimeter modes are soft clamped due to symmetry and the boundary conditions at the polygon vertices. Perimeter modes reach $Q$ of 3.6 billion at room temperature while spanning only two acoustic wavelengths---a 4-fold improvement over the state-of-the-art mechanical $Q$ with 10-fold smaller devices. The small size of our devices makes them well-suited for near-field integration with microcavities for quantum optomechanical experiments. Moreover, their compactness allows the realization of phononic lattices. We demonstrate a one-dimensional Su-Schrieffer-Heeger chain of high-$Q$ perimeter modes coupled via nearest-neighbour interaction and characterize the localized edge modes.
	
\end{abstract}
\maketitle

High-$Q$ mechanical resonators have found widespread application in precision measurements, including interferometric detection of gravitational waves \cite{gonzalez1994brownian} and optomechanical experiments exploring the limits of quantum measurement \cite{rossi2018measurementbased,wilson2015measurementbased,mason2019continuous}. In these endeavors, mechanical dissipation plays a central role as it describes the coupling rate to the thermal bath, and together with the resonator’s effective mass sets the thermal force noise floor \cite{aspelmeyer2014cavity,sementilli2021nanomechanical}. High force sensitivity requires low mass and low dissipation and one typically comes at the cost of the other, as mechanical losses are often dominated by surface effects, and it has therefore been observed that smaller mechanical resonators have higher dissipation \cite{villanueva2014evidence}.

Over the last decade, the experimentally accessible quality factors of nanomechanical resonators have increased by three orders of magnitude. This revolution in dissipation control is underpinned by the phenomenon ‘dissipation dilution’, first seen in test mass suspensions of gravitational wave detectors \cite{gonzalez1994brownian}. Dissipation dilution occurs through a combination of stress and geometric nonlinearity \cite{fedorov2019generalized}; the condition where resonator elongation is quadratic in mode amplitude. Notably, dissipation dilution allows reduction of dissipation by engineering the geometry of the resonator rather than its constituent materials. It was shown that in the regime of strong dissipation dilution, most of the mechanical losses of an unpatterned structure arises from mode curvature near the boundaries \cite{yu2012control,tsaturyan2017ultracoherent}. A number of techniques were therefore developed which reduce the mechanical mode amplitude near the clamping points, including phononic bandgap engineering \cite{tsaturyan2017ultracoherent}, clamp tapering \cite{bereyhi2019clamp} and hierarchical structuring \cite{beccari2021hierarchical}. The elimination of these boundary losses (soft clamping) has allowed nanomechanical resonators to approach quality factors of one billion at room temperature \cite{ghadimi2018elastic,beccari2021hierarchical,beccari2021strained}---equaling the best macroscopic resonators.

To realize high mechanical quality factors by dissipation dilution, extremely high aspect ratios (i.e. length/thickness $> 10^5$) are required: state-of-the-art implementations have thicknesses of tens of nanometers and lengths on the millimeter scale \cite{reinhardt2016ultralownoise,tsaturyan2017ultracoherent,ghadimi2018elastic,beccari2021hierarchical}. For example, phononic bandgap engineering typically requires a structure size of at least ten acoustic wavelengths. These large sizes present an obstacle to the practical use of such resonators, especially in integrated optomechanical systems, where the mechanical resonator is suspended within hundreds of nanometers from an optical microcavity to engineer strong near-field optomechanical coupling \cite{schilling2016near,guo2019feedback}. The large size requirement also limits the attainable quality factors in such resonators due to the difficulty of suspending stressed resonators larger than a few millimeters.

\begin{figure*}[t]
	\includegraphics[width = \linewidth]{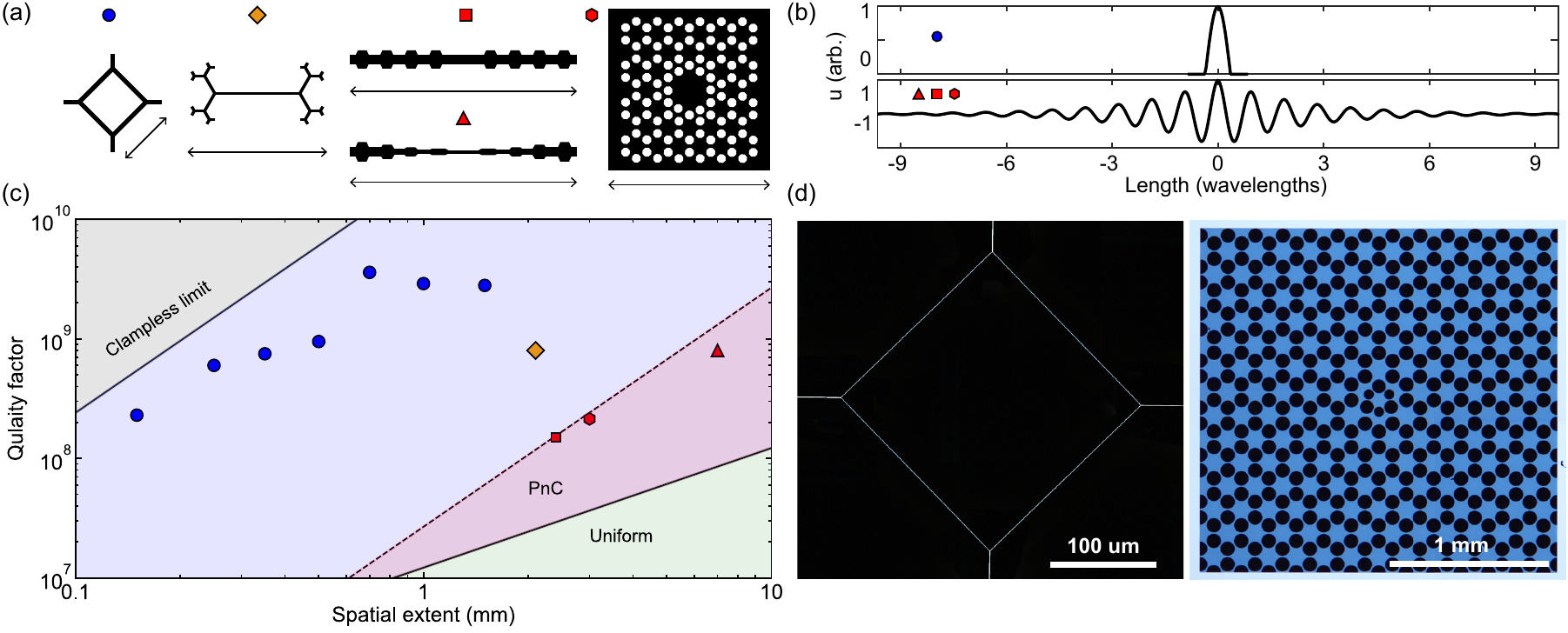}
	\caption{\textbf{Comparison of state-of-the-art strained mechanical resonator designs.} (a) Schematics of different mechanical resonators with their spatial extent illustrated by arrows. From left to right: polygon resonator, binary tree nanobeam \cite{beccari2021hierarchical}, phononic crystal nanobeam \cite{ghadimi2018elastic}, strain engineered nanobeam \cite{ghadimi2018elastic} and phononic crystal membrane \cite{tsaturyan2017ultracoherent}. (b) Displacement profile of perimeter modes (plotted between two adjacent clamps) and phononic crystal soft-clamped modes with the same frequency. (c) Measured room temperature $Q$ for polygons with different lengths (blue circles) and the best reported values for other designs of tensioned resonators \cite{beccari2021hierarchical,ghadimi2018elastic,tsaturyan2017ultracoherent}. The green shaded area is accessible for the fundamental mode of a tensioned nanobeam. The red shaded region is accessible by the phononic crystal designs shown in panel a. The dashed line shows the soft clamping limit for the $30^\mathrm{th}$ order mode and the solid black line separating the blue and gray region is the soft clamping limit for the fundamental mode. (d) Size comparison between a $\SI{200}{\micro\meter}$ long polygon resonator with $Q=155\times 10^6$ at $\SI{1.6}{\mega\hertz}$ and a $\SI{2}{\milli\meter} \times \SI{2}{\milli\meter}$ phononic crystal membrane with $Q = 74\times 10^6$ at $\SI{1.46}{\mega\hertz}$ frequency \cite{fedorov2020thermal}, illustrating the reduced size.}
	\label{fig:intro}
\end{figure*}

 Our best resonators show quality factors exceeding 3 billion at room temperature, exceeding the state of the art by a factor of four in 10-fold smaller devices (\fref{fig:intro}{(c)}). These structures are well-suited for integration with nanoscale cavities for sensing and quantum optomechanical experiments. The simplicity of the design further increases its flexibility: by scaling the size of the resonator, soft-clamped perimeter modes can be realized at different frequencies. In our work, we show $Q>10^8$ over a span of more than 4 octaves (\SI{170}{\kilo\hertz} to \SI{2.5}{\mega\hertz}), but smaller and larger frequencies are also possible.
The perimeter modes of different polygon resonators can be controllably coupled through a joint tether, allowing the creation of phononic dimers. Phononic dimers have been studied in defect modes of phononic crystals \cite{Xiaochun2005coupling, Miyashita2008analysis, Lanzillotti2010phononic, Khelif2003experimental}, targeting force sensing applications \cite{catalini2020softclamped}. The compactness of the resonators also simplify the creation of large, coupled phononic arrays, making polygon resonators a suitable platform for exploring multimode physics of phononic structures, such as topological modes \cite{Susstrunk2015observation,  Nash2015topological, Peano2015topological, Lu2016observation, Miniaci2018experimental, Schmidt2015optomechanical}. As a proof-of principle, we demonstrate a one-dimensional array of six resonators forming a Su-Schrieffer-Heeger (SSH) chain and characterize the localized edge modes.

\section{Soft-clamped perimeter modes}

\begin{figure*}[t]
	\includegraphics[width = \linewidth]{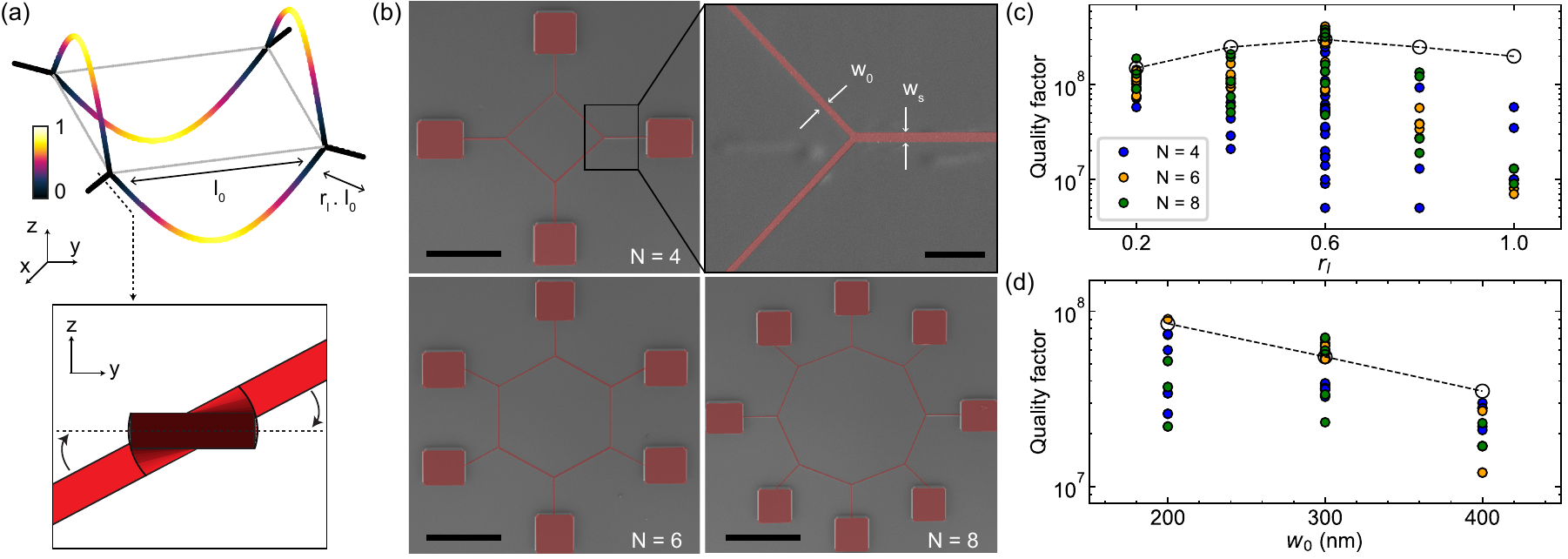}
	\caption{\textbf{Ultra-high-$Q$ perimeter modes.} (a) FEM simulations of the perimeter mode with an inset showing the generation of torsional deformation of the tethers. (b) False-colored scanning electron micrographs of polygon \SiN\ strained resonator devices with $l_0 = \SI{700}{\micro\meter}$, $r_l = 0.6$, $w_0 = \SI{200}{\nano\meter}$ and $\SI{20}{\nano\meter}$ thickness (scale bars correspond to $\SI{500}{\micro\meter}$). Top right of b shows a zoomed-in junction of a stress-preserving square resonator (scale bar corresponds to $\SI{1.5}{\micro\meter}$). (c) The effect of tether length ($r_l$) on $Q$ for samples with $l_0 = \SI{250}{\micro\meter}$, $w_0 = \SI{300}{\nano\meter}$ and $\SI{20}{\nano\meter}$ thickness. Blue, red and orange data points represent square, hexagon and octagon resonators respectively. Open circles (joined by a dashed line) are FEM simulations. (d) The effect of polygon width ($w_0$) on the $Q$ for samples with $l_0 = \SI{100}{\micro\meter}$, $r_l = 0.6$ and $\SI{20}{\nano\meter}$ thickness).}
	\label{fig:QvsGeometry}
\end{figure*}

For a resonator experiencing dissipation dilution, its quality factor is enhanced by the dissipation dilution factor ($D_Q$) \cite{fedorov2019generalized} over the intrinsic quality factor of the material ($Q_{\mathrm{int}}$) such that its quality factor can be found as $Q=D_Q\times Q_{\mathrm{int}}$. The intrinsic quality factor is a material property, given by $Q_{\mathrm{int}}=1/\phi$ where $\phi$ is the loss angle characterizing the delay between stress and strain. However, $D_Q$ can be engineered by changing the resonator geometry, and here we show how perimeter modes have more than two orders of magnitude increased $D_Q$ compared to uniform strings.

Our resonators can be understood as a network of tensioned one-dimensional strings. It is therefore useful to start from the dilution factor of the $n$th flexural mode of a stressed mechanical resonator \cite{fedorov2019generalized}: 
\begin{equation}\label{eq: D_qn}
	D_{Q,n} = \frac{1}{\alpha_n \lambda + \beta_n \lambda^2},\ \lambda = \sqrt{\frac{1}{12\epsilon}}\frac{h}{l}
\end{equation}
\noindent with thickness $h$, characteristic length $l$ and average strain $\epsilon$. The $\alpha_n\lambda$ term arises from the curvature near the boundaries and $\beta_n \lambda^2$ from the curvature distributed over the rest of the mode. For structures with high tension and large aspect ratio, $\lambda$ is much smaller than one, and $\alpha_n $ therefore sets the limit on the dilution factor. The boundary curvature ($\alpha_n$) can be suppressed by soft-clamping techniques, two of which were previously reported: phononic bandgap engineering \cite{tsaturyan2017ultracoherent,ghadimi2018elastic} and hierarchical structuring \cite{beccari2021hierarchical}. Here, we show that a different class of resonators also host soft-clamped modes: polygons composed of connected beams that are tethered at their vertices. \fref{fig:QvsGeometry}{(a)} shows an example polygon resonator with four vertices and the simulated displacement profile of its fundamental perimeter mode, i.e. the mode which has half an acoustic wavelength per polygon side. \fref{fig:QvsGeometry}{(b)} shows fabricated polygon resonators with four, six and eight vertices.

The dissipation dilution of perimeter modes of polygon resonators can be found analytically under the narrow-beam approximation, in which the length over width ratios of all constituent beams are infinitely large. In this case, the boundary loss coefficient ($\alpha_n$) is zero, since perimeter modes do not produce out-of-plane displacements in the tethers and therefore never reach clamped boundaries. The torsional deformations of the tethers, however, are always non-negligible, and they add to the distributed loss coefficient $\beta_n$. The dilution factor equals the ratio of the tension energy stored by the sinusoidal standing waves in the perimeter beams to the sum of the bending energy of those waves and the torsional energy of the tethers. For an $N$-sided polygon with equal stress along all segments $D_Q^{-1}$ is found as
 
\begin{equation}\label{eq: D_q_theory}
	D_{Q}^{-1} = \left(\frac{1}{n^2 \pi^2 \lambda^2 }\right)^{-1} + \left(\frac{r_l(1+\nu)\cos^2{(\pi/N)}}{4r_w \lambda^2}\right)^{-1}
\end{equation}

\noindent where $r_l$ is the ratio of the support length to the side length of the polygon (\fref{fig:QvsGeometry}{(a)}), $r_w = w_s/w_0$ is the ratio of the support width to the side width of the polygon, $n$ is the perimeter mode order and $\nu$ is the Poisson ratio of the material. The predictions of \eqref{eq: D_q_theory} are in good quantitative agreement with experimental results for a subset of devices in our work (the narrowest devices). For the others, it still provides qualitative insights, but we resort to two-dimensional finite-element simulations for quantitative theoretical predictions. The validity range of \eqref{eq: D_q_theory} is investigated in detail in Appendix \ref{appendix: theory}. One qualitatively new effect introduced by finite aspect ratio is a finite boundary loss coefficient ($\alpha_n$), which in real structures is suppressed by a factor of order $(w_s/l_s)^2$, where $l_s$ is the length and $w_s$ is the width of the supports. The $Q$ width dependence observed in FEM simulations is also not captured by this simple model, as it arises due to the torsional deformation of the polygon sides. The dilution factor of the perimeter modes shows a $\lambda^{-2}$-scaling, which is a signature of soft clamping also achieved in fundamental modes of binary tree nanobeams \cite{beccari2021hierarchical} and localized modes of phononic crystal nanobeams \cite{ghadimi2018elastic} and membranes \cite{tsaturyan2017ultracoherent}.

\begin{figure*}[t]
	\includegraphics[width = \linewidth]{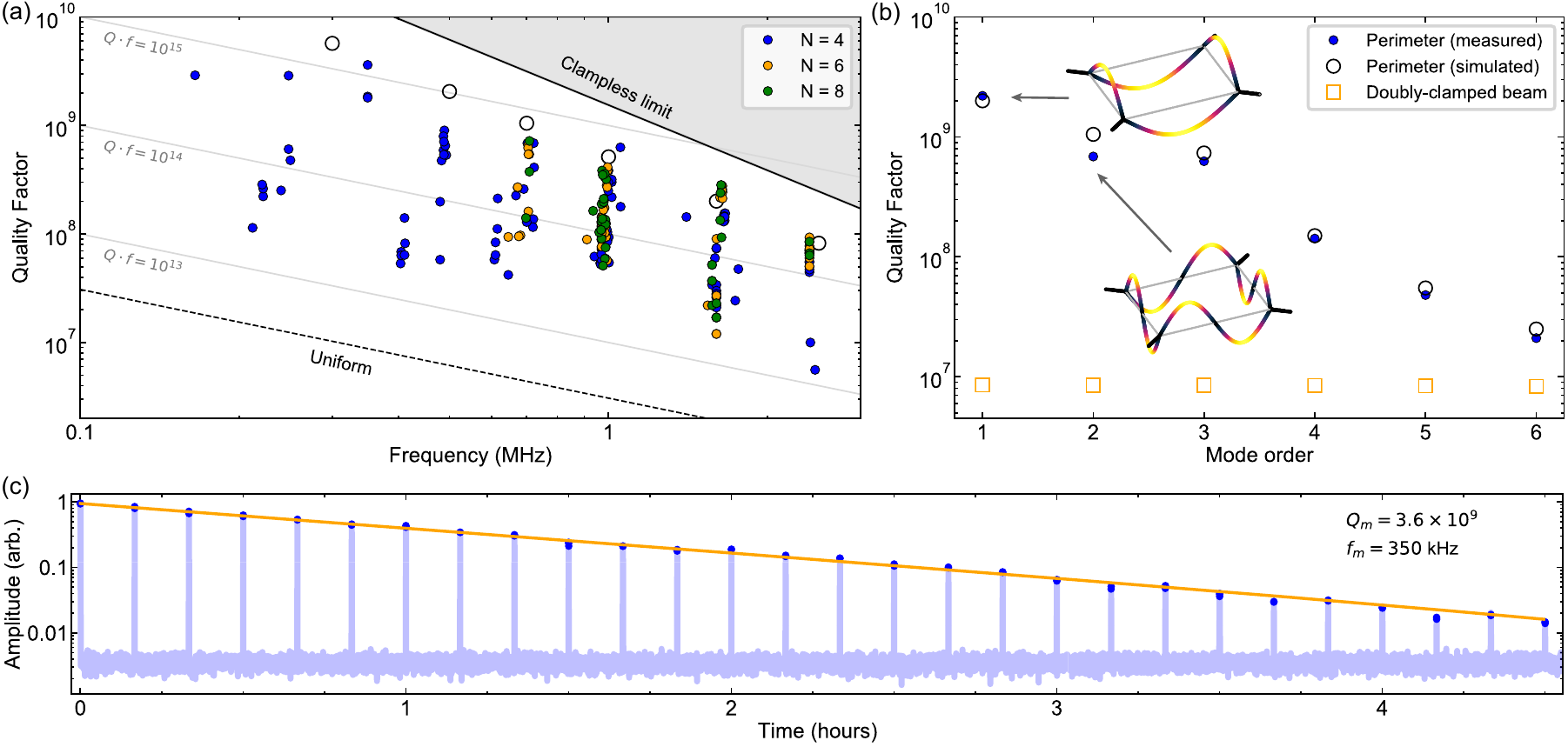}
	\caption{\textbf{Characterization of fundamental and higher-order perimeter modes.} (a) $Q$s of fundamental perimeter modes for square, hexagon and octagon-shaped \SiN\ resonators (blue, orange and green circles respectively) with $r_l \in [0.2, 0.4, 0.6, 0.8]$ and $w_0 = \SI{200}{\nano\meter}$ and $\SI{300}{\nano\meter}$. FEM simulation $Q$s (open circles) are plotted for optimal geometrical parameters ($r_l=0.6$ and $w_0=\SI{200}{\nano\meter}$). The solid black line shows the clampless limit. Dashed black line: fundamental mode $Q$  of a uniform nanobeam with the same mode frequency. (b) High-order perimeter mode $Q$s (blue dots) of a square with $l_0 = \SI{700}{\micro\meter}$, $r_l = 0.2$ and $w_0 = \SI{200}{\nano\meter}$ is shown with FEM predictions (open circles). Open squares show the calculated $Q$s of a uniform beam with the same fundamental mode frequency. Displacement profiles of the first two perimeter modes are shown as insets. (c) Gated ringdown measurement (\SI{1}{\second} on, \SI{10}{\minute} off) of a square with $l_0=\SI{700}{\micro\meter}$, $r_l=0.6$ and $w_0 = \SI{200}{\nano\meter}$ with a fundamental perimeter mode $Q$ of $3.6 \times 10^9$ at $\SI{350}{\kilo\hertz}$. Orange line: exponential fit to the data shown by blue circles. Faded blue data are excluded from the fit as the laser is blocked at these times.}
	\label{fig:Qall}
\end{figure*}

 \section{Experimental characterization of perimeter modes}
 
We fabricate devices out of $\SI{20}{\nano\meter}$-thick \SiN\ film with $\SI{1}{\giga\pascal}$ deposition stress, following a previously established fabrication process focused on creating large gaps between the suspended resonators and the substrate \cite{ghadimi2018elastic,Groth_Bereyhi_2020} (see Appendix \ref{appendix: fab}). The intrinsic quality factor of the film has been previously characterized to be $Q_{\mathrm{int}}=2600$ \cite{beccari2021hierarchical} and we use this value in all FEM simulations. Mechanical $Q$s are extracted from ringdown decay times of excited modes that are recorded using a free-space optical interferometer with gated laser illumination of the sample (see ref. \cite{beccari2021hierarchical} for more details). An example ringdown trace is shown in \fref{fig:Qall}{(c)}. The chips are rested on a holder inside the vacuum chamber without any fixation unless otherwise indicated. We investigate the effect of chip mounting on the $Q$ in detail in Appendix \ref{appendix: chip_mount}, where we find that perimeter modes are less sensitive to chip clamping conditions than the low order modes of unpatterned nanobeams \cite{schmid2011damping} and membranes \cite{wilson2009cavity}. 

We fabricate and characterize devices with different values of relative support length ($r_l$) and segment width. We expect an initial increase in $Q$ as a function of support length due to the reduction of lossy torsional energy. However, as $r_l$ reaches one, torsional modes of the supports start to hybridize with the perimeter modes leading to a dramatic increase in dissipation (see Appendix \ref{appendix: theory}). This is reflected in FEM simulations and measurements, with a decrease in $Q$ starting from $r_l=0.8$ (see \fref{fig:QvsGeometry}{(c)}). We next investigate the influence of overall resonator width by fabricating and characterizing polygons with varying widths and $l_0 = \SI{100}{\micro\meter}$ and $r_l = 0.6$ (the optimal $r_l$ value found in \fref{fig:QvsGeometry}{(c)}). We observe an increase in the $Q$ by decreasing the resonator's width $w_0$, which is in good agreement with simulation.

\begin{figure*}[t]
	\includegraphics[width = \linewidth]{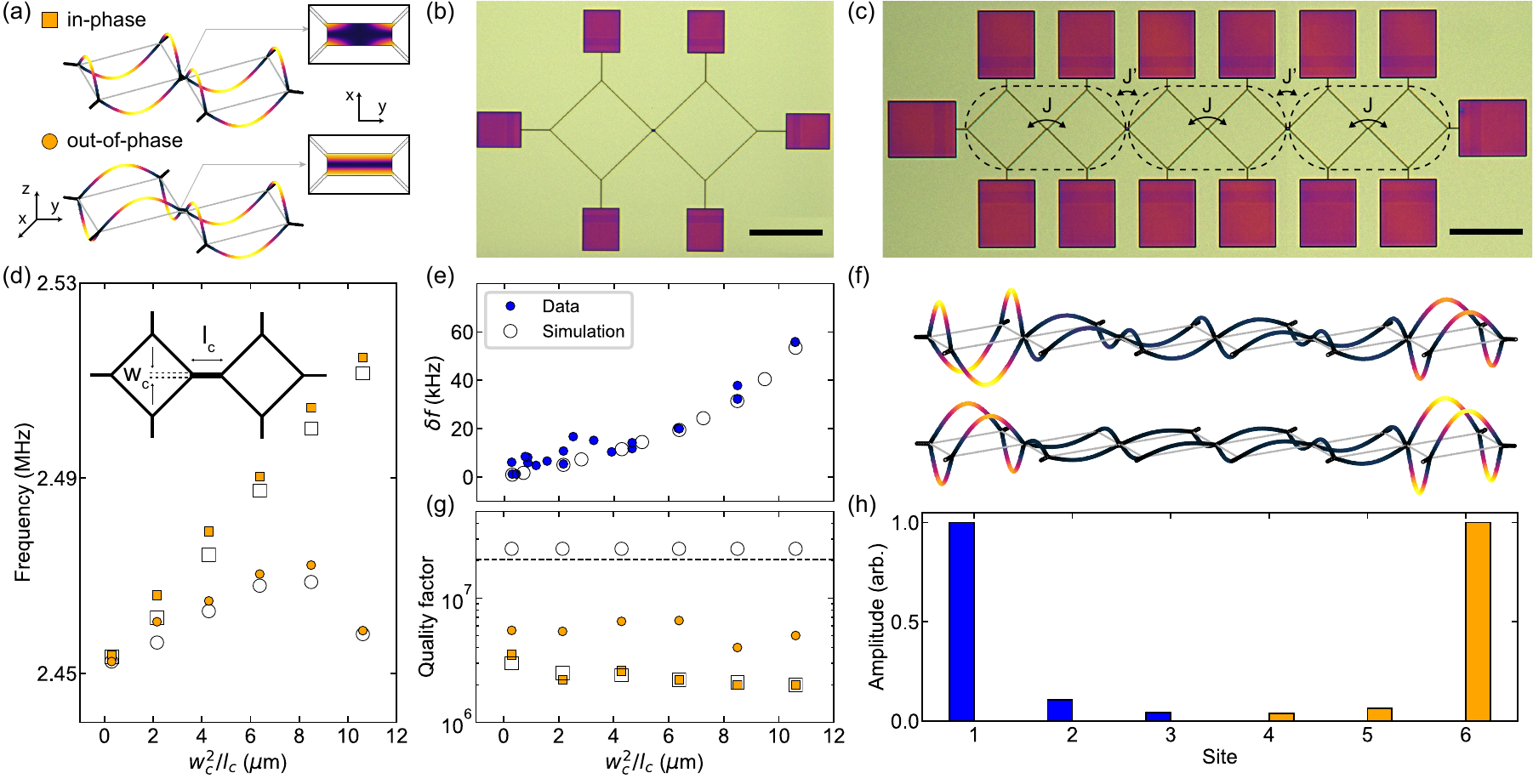}
	\caption{\textbf{Phononic dimers and arrays of polygon resonators.} (a) FEM simulations of the in-phase and out-of-phase modes of a phononic dimer. The insets show the deformation of the coupler segment for the two cases. (b, c) Optical micrographs of a fabricated dimer and a chain; both composed of square-shaped resonators with side lengths of $\SI{100}{\micro\meter}$. The scale bars correspond to $\SI{100}{\micro\meter}$. (d) The frequencies of the in-phase (square) and out-of-phase (circle markers) modes from data (filled markers) and FEM simulation (empty markers) for different coupling parameters. The inset indicates the coupling parameters of a phononic dimer. (e) Frequency splitting of the in-phase and the out-of-phase modes from data (filled markers) and FEM simulation (empty markers) for different coupling parameters. (f) FEM simulation of the edge mode profiles of a phononic array made of six square-shaped resonators. The slight asymmetry is due to the disorder in the mesh of the simulation. (g) $Q$ of the hybridized modes from data (filled markers) and FEM simulation (empty markers) for different coupling parameters. The dashed line corresponds to quality factor of a single square-shaped resonator.  (h) RMS values of the Brownian motion of each site measured for the two split modes, showing localization of the mode at the edges of the chain.}
	\label{fig:Dimer_Chain}
\end{figure*}

In \fref{fig:Qall}{(a)}, we survey the measured quality factors of fundamental perimeter modes in resonators with different geometries: $N=4,6$ and 8, $r_l = 0.2, 0.4, 0.6$ and $0.8$, and $w_0 = \SI{200}{\nano\meter}$ and $\SI{300}{\nano\meter}$. Our highest measured $Q$ is $3.6\times 10^9$ for a $\SI{700}{\micro\meter}$ square with a fundamental perimeter mode at $\SI{350}{\kilo\hertz}$. \fref{fig:Qall}{(c)} shows a sample ringdown measurement of our best device with one second gates separated by \SI{10}{\minute}. For each mode, we perform multiple measurements with varying duty cycle to rule out optical damping or anti-damping effects on the $Q$ for each sample. The highest quality factors we observe are in good agreement with the FEM simulations. In \fref{fig:Qall}{(a)}, simulation values are calculated for each length for the optimal fabricated design ($w_0 = \SI{200}{\nano\meter}$, $r_l=0.6$). The clampless limit shown in \fref{fig:Qall}{(a)} is the $Q$ of a mode of the same frequency with zero boundary losses ($\alpha_n=0$) and thereby a dilution factor given by the first term of \eqref{eq: D_q_theory}. Simulated $Q$s of the perimeter modes for the optimum design are about $2/3$ of the clampless limit with the same scaling of $D_{Q,1} \propto 1/\lambda^2$. This scaling shows that the perimeter modes are soft-clamped and that the boundary losses are negligible. The discrepancy from the clampless limit arises due to torsion of the tethers.

In polygon resonators, the boundary losses are eliminated for a whole family of perimeter modes that have nodes at the tethers. We characterize the $Q$ of the first 6 harmonics of the perimeter mode for a square with $l_0=\SI{700}{\micro\meter}$, $w_0=\SI{200}{\nano\meter}$ and $r_l=0.2$. As expected from simulations, the harmonics of the perimeter mode also show significantly higher $D_Q$ with respect to a doubly-clamped beam (\fref{fig:Qall}{(b)}).

\section{Phononic dimers and arrays}

The torsional deformations in the tethers can be utilized to couple the perimeter modes of two polygon resonators (\fref{fig:QvsGeometry}{(a)}).
By joining two square-shaped resonators with a tether, we induce torsional coupling between the perimeter modes (as in the device in \fref{fig:Dimer_Chain}{(b)}).
This coupling results in hybridization of the two perimeter modes of the two resonators into a pair of new modes. As shown in \fref{fig:Dimer_Chain}{(a)}, these modes are `in-phase' and `out-of-phase' combinations of the two perimeter modes. Due to the symmetries of these modes, the deformation of the coupler segment takes two different forms. For the in-phase mode, the anti-symmetric displacement results in torsional deformation while for the out-of-phase mode the deformation is solely rotational and therefore produces no elastic energy (\fref{fig:Dimer_Chain}{(a)}).
The strength of the coupling (i.e. the hybridized modes' frequency splitting) is determined by the dimensions of the coupler segment ($w_c$ and $l_c$, shown in the inset of \fref{fig:Dimer_Chain}{(d)}).
Using FEM simulations, we find an approximately linear relation between the frequency splitting, $\delta f$, and the parameter $w_c^2/l_c$ (see Appendix \ref{sec:SI_dimers_simulations} for more details).
We experimentally study `phononic dimers' with side lengths of $\SI{100}{\micro\meter}$  with different coupling parameters $w_c^2/l_c$ and find the expected hybridization of perimeter modes (see \fref{fig:Dimer_Chain}{}).
In \fref{fig:Dimer_Chain}{(e)}, we show the scaling of $\delta f$ with the coupling parameter, which for strong coupling are in agreement with the FEM simulations. For weaker coupling ($\delta f<\SI{10}{\kilo\hertz}$), the fabrication disorder can be larger than the coupling, which prevents hybridization (see Appendix \ref{sec:SI_dimers_simulations} for details). 

The difference in the deformation of the coupling segments also leads to different dissipation rates for the hybridized modes. The extra torsion of the in-phase mode results in lower quality factors for these modes (as shown experimentally in \fref{fig:Dimer_Chain}{(g)}). In contrast, the out-of-phase mode is missing one torsional loss contribution compared to a single square-shaped resonator, and therefore has a slightly higher quality factor. This is supported by FEM simulations, and while the experimental quality factors of the out-of-phase modes do not attain the predicted values, they do show reduced loss compared to the in-phase modes (\fref{fig:Dimer_Chain}{(g)}). 

 The same coupling principle can be used in arrays of coupled polygon resonators, where the controllable coupling rates allow the engineering of phononic modes of the array. By choosing coupling rates that alternate between intra and inter cell coupling, $J$ and $J'$, one can realize a one-dimensional Su-Schrieffer-Heeger (SSH) chain \cite{Su1979solition}. \fref{fig:Dimer_Chain}{(c)} shows a fabricated array of 6 square resonators with side length of $\SI{100}{\micro\meter}$ and alternating coupling rates with simulation values of $J/2\pi=\SI{0.35}{\kilo\hertz}$ and $J'/2\pi=\SI{5}{\kilo\hertz}$ (corresponding to dimer frequency splittings of $\SI{0.7}{\kilo\hertz}$ and $\SI{10}{\kilo\hertz}$). In this case the array can be seen as a three-cell chain of phononic dimers. If the inter-cell coupling is weaker than the intra-cell coupling ($J' > J$), the array exhibits edge modes of the perimeter modes of the chain. \fref{fig:Dimer_Chain}{(f)} shows FEM simulations of such a mode of this array. This is a collective mode where all the resonators are oscillating in their perimeter mode while the amplitudes of the modes at the `edge' sites of the array are much greater than for the `bulk' sites. However, in practice, fabrication disorder can lift the degeneracy and the two new modes which emerge are not fully symmetric but nevertheless show some edge mode characteristics. We observe two such modes in our structure and characterize them by measuring the RMS values of the thermal fluctuations of each mode at each site and observe that both modes are fully localized at one of the edge sites (see \fref{fig:Dimer_Chain}{(h)}). In an alternative scenario where $J'<J$, the system transitions into a so-called `trivial' phase where the edge modes' degeneracy is fully lifted and their edge characteristics completely vanish. We fabricate an array with flipped order of strong-weak coupling rates ($J/2\pi=\SI{5}{\kilo\hertz}$ and $J'/2\pi=\SI{0.35}{\kilo\hertz}$) and follow the same measurement procedure. We identify six modes that have frequencies close to that of a single perimeter mode but we do not observe any edge structure in any of them (see Appendix \ref{sec:SI_arrays_full_spectra} for the complete mode spectrum). This provides further evidence for topological symmetries playing a significant role in the eigenmodes of our arrays. We measure $Q = 10^7$ for the edge modes of the SSH chain which is $50\%$ below the simulation. The simulated $Q$ factor for the edge modes is equal to the single cell perimeter mode $Q$ factor, showing that the chains of polygon resonators are capable of retaining the high quality factors.
 
\section{Conclusion}
We demonstrated extremely compact perimeter modes in vertex-clamped polygon resonators that implement soft clamping without a phononic bandgap. Our devices have $Q$s as high as $3.6\times 10^9$ at room temperature, exceeding the state of the art by a factor of four with ten times smaller devices. In addition, we show coupling of polygon resonators forming phononic dimers with controllable mechanical mode splitting which allows us to explore coupled high-$Q$ phononic arrays. Coupled polygon resonators could open new opportunities for studying topological properties of nanomechanical resonators in one and two dimensional arrays.

The compact size of the polygon resonators is crucial for monolithic integration in the near-field of optical cavities  \cite{schilling2016near,gavartin2012hybrid,guo2019feedback}. Near-field coupling allows exceptionally high optomechanical coupling rates, but requires small gaps (order of \SI{100}{\nano\meter}) between the optical cavity and mechanical resonator, making it technically challenging with millimeter-scale structures. The integration of high $Q$ polygon resonators with optical cavities such as photonic crystals, would be especially advantageous for room temperature quantum optomechanical experiments such as quantum feedback\cite{wilson2015measurementbased,rossi2018measurementbased}, ponderomotive squeezing \cite{safavi2013squeezed,purdy2013strong,sudhir2017appearance,aggarwal2020room} and observation of radiation pressure shot noise \cite{purdy2013observation,cripe2019measurement}. Efficient interferometric readout of such low dissipation mechanical resonators could also allow cavity-free feedback cooling of the mechanical motion to the quantum ground state \cite{poggio2007feedback,pluchar2020towards,tebbenjohanns_quantum_2021}.

Owing to the record-high $Q$, small size and ultra low dissipation rates, our highest $Q$ resonator allows thermal noise-limited force sensitivity of $\sqrt{S_F} = \sqrt{4k_BTm\Gamma_m}\approx \SI{413}{zN/\sqrt{\hertz}}$ at room temperature, showing good prospects for force sensing applications. In addition, polygons are very well-suited for integration in sensing platforms, such as inverted microscope systems, as the multiple sides facilitate addition of several probes and measurement specimens on the same mechanical resonator \cite{gruber2019mass,yang2006zeptogram,halg2021membranebased,kovsata2020spin}. Moreover, the coupled polygons provide an alternative platform for parametric spin sensing owing to their controllable coupling rates and potential for extremely low dissipation \cite{catalini2020softclamped,kovsata2020spin}.
\\
\section*{Acknowledgments}

This work was supported by funding from the Swiss National Science Foundation under grant agreement no. 182103, the EU H2020 research and innovation programme under grant agreement no.732894 (HOT) and the European Research Council grant no. 835329 (ExCOM-cCEO). G.H. and N.J.E. acknowledge support from the Swiss National Science Foundation under grant no. 185870 (Ambizione). All samples were fabricated at the Center of MicroNano Technology (CMi) at EPFL. Data and data analysis codes are available at \texttt{Zenodo} upon publication.


\appendix
\section{Theory model for dissipation dilution of the perimeter modes}\label{appendix: theory}
To analytically estimate the dilution factor of perimeter modes, we proceed as in \cite{yu2012control,fedorov2020fractallike}. We express a lossless ``tension" energy:
\begin{equation}
    \langle W_\mathrm{tens}\rangle = N\sigma w_0 h \int_0^{l_0} \left(u'(x)\right)^2 dx,
    \label{eqn:tens_energy}
\end{equation}

\noindent from the out-of-plane displacement profile $u(x)$ along the polygon sides. Here, $N$ is the number of sides, $\sigma$ is the static stress after relaxation, $w_0$ is the polygon width, $l_0$ the side length and $h$ the film thickness.

The lossy energy is approximated as arising only from bending in the polygon sides and from the torsion of the supporting tethers \cite{fedorov2020fractallike}. Explicitly,

\begin{equation}
    \langle W_\mathrm{bend}\rangle = \frac{N E w_0 h^3}{12} \int_0^{l_0} \left(u''(x)\right)^2 dx,
    \label{eqn:bend_energy}
\end{equation}

\noindent with Young's modulus $E$ and the integral running over the polygon side, and

\begin{equation}
    \langle W_\mathrm{tors}\rangle = \frac{N E w_s h^3}{6\left(1+\nu\right)} \int_0^{l_s} \left(\tau'(x)\right)^2 dx,
    \label{eqn:torsion_energy}
\end{equation}

\noindent with $\tau$ torsion angle, evaluated over the supporting tether of length $l_s$ and width $w_s$. Note that we are neglecting torsion in the polygon side beams, which can contribute significantly for larger side widths.

In order to evaluate the integrals, we assume simplified displacement and torsion profiles. In the polygon sides, the out-of-plane displacement will be almost perfectly sinusoidal for $\lambda \ll 1$, with nodes at the polygon vertices: $u(x) = u_0\,\mathrm{sin}\left(n\pi x/l_0\right)$ ($n$ is the perimeter mode index). The torsion angle is assumed to decay linearly towards the clamping points \cite{fedorov2020fractallike}. This approximation is acceptable when the support tether length is much smaller than the torsional wavelength at the mechanical frequency of the perimeter mode. Torsion and out-of-plane displacement are connected by continuity of the displacement field at the joint:

\begin{equation}
    \tau\left(l_s\right) = \frac{u'\left(0\right)}{\mathrm{sin}\left(\theta\right)} = \frac{n\pi u_0}{l_0\cdot\mathrm{cos}\left(\pi/N\right)},
\end{equation}

\begin{figure}[b]
    \centering
    \includegraphics[width=\linewidth]{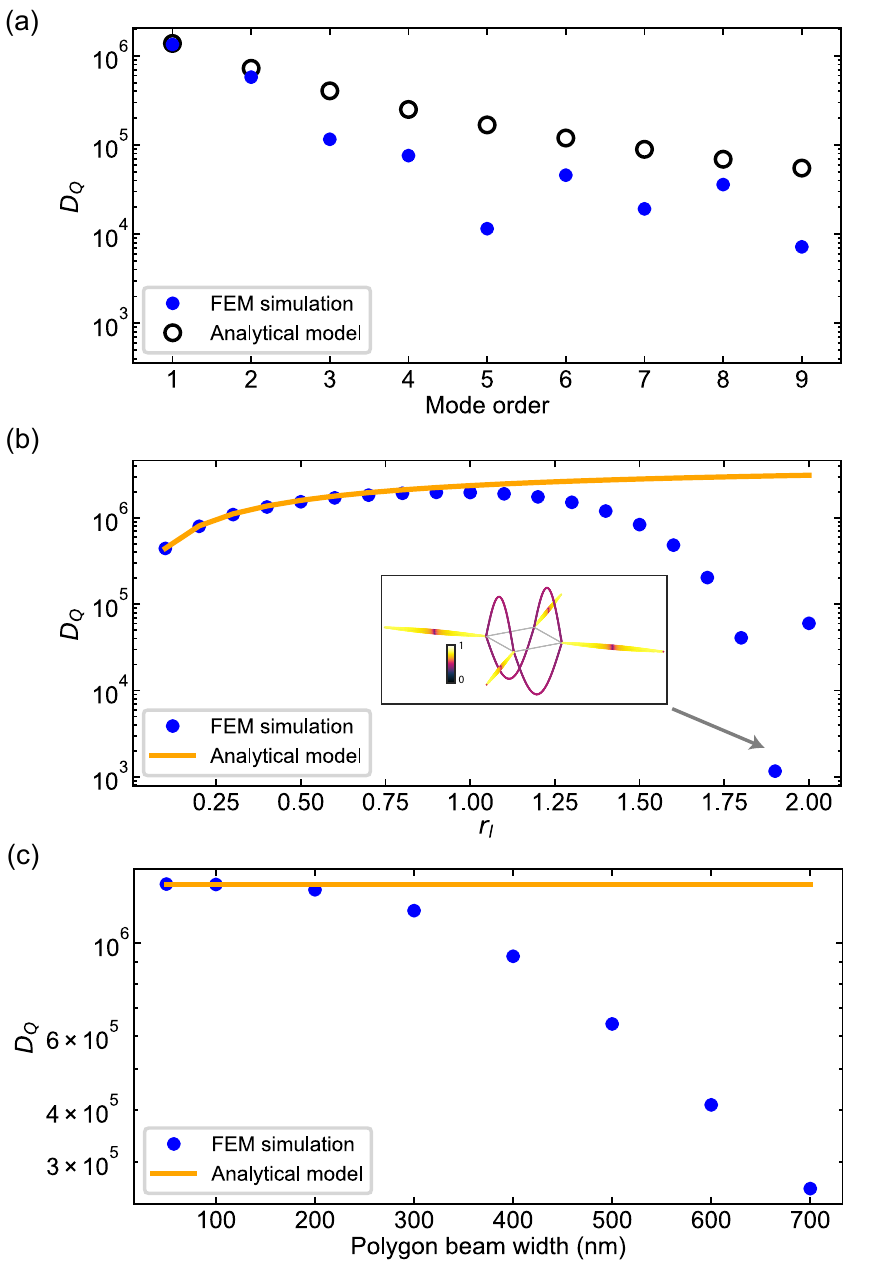}
    \caption{\textbf{Theory model and FEM simulation comparison.} We compare simulation and analytical predictions for a polygon resonator with $N = 4$, $l_0 = \SI{700}{\mu m}$ and uniform stress ($r_w = \sqrt{2}$). (a) Dilution factor ($D_Q$) for different perimeter modes ($w_0 = \SI{200}{nm}$, $r_l = 0.4$). (b) $D_Q$ of the first perimeter mode as the support length is varied (fixed $w_0 = \SI{200}{nm}$). The inset is the mode displacement for $r=1.9$, with the color encoding the torsional energy density. (c) $D_Q$ of the first perimeter mode as the width of all the segments is varied (fixed $r_l = 0.4$)}
    \label{fig_SI:Theory}
\end{figure}

\noindent where $\theta = \pi/2 - \pi/N$ is the semi-angle subtended by two polygon sides.

Since $D_Q = \langle W_\mathrm{tens}\rangle/\left(\langle W_\mathrm{bend}\rangle + \langle W_\mathrm{tors}\rangle\right)$, we can distinguish two separate contributions:

\begin{equation}
    D_\mathrm{Q,bend} = \frac{1}{n^2\pi^2\lambda^2},
\end{equation}

\begin{equation}\label{eq_SI:W_tor}
    D_\mathrm{Q,tors} = \frac{r_l\left(1+\nu\right)\mathrm{cos}^2\left(\pi/N\right)}{4\left(r_w\right)\lambda^2},
\end{equation}

\noindent such that $D_Q^{-1} = D_\mathrm{Q,bend}^{-1} + D_\mathrm{Q,tors}^{-1}$. Here, we defined $r_l = l_s/l_0$, $r_w = w_s/w_0$ and $\lambda = \sqrt{E/\left(12\sigma\right)}h/l$.

The static stress in the polygon segments $\sigma$ can be found by imposing force balance and conservation of the length of the path separating two polygon clamping points, upon structure release:

\begin{equation}\label{eq:sigma_SI}
    \sigma = \frac{\sigma_0\left(1-\nu\right)\left(1+2r_l\cdot\mathrm{sin}\left(\pi/N\right)\right)}{1+4r_l/r_w\cdot\mathrm{sin}^2\left(\pi/N\right)},
\end{equation}

\noindent where $\sigma_0$ is the thin film deposition stress, prior to structure undercut.

Our model provides a good approximation of the $D_Q$ of the first perimeter mode, but in general overestimates it, as evident from the comparison with FEM simulations (\fref{fig_SI:Theory}{}). This is mainly due to the simplified assumption that lossy elastic energy is only stored in tether torsion and polygon segments out-of-plane bending, which is valid in the limit where $w/l\rightarrow 0$. In the real displacement patterns, the torsion in the polygon segments is also significant, and becomes dominant as the segments get wider (see \fref{fig_SI:Theory}{(c)}). Moreover, the model is less accurate for increasing mechanical frequency and tether length, as the torsion angle profile in the tethers can no longer be approximated as linear (see \fref{fig_SI:Theory}{(a,b)}). In \fref{fig_SI:Theory}{(b)}, a minimum of $D_Q$ is observed as the first perimeter mode hybridizes with a tether torsional mode, around $r \approx 1.9$. As a broad criterion, the analytical expression is accurate for the first perimeter mode, and $r_l \leq 0.6$, $w_0 \leq \SI{200}{nm}$.

\section{Effect of chip mounting on the $Q$}\label{appendix: chip_mount}
\begin{figure}[h]
    \centering
    \includegraphics[width=\linewidth]{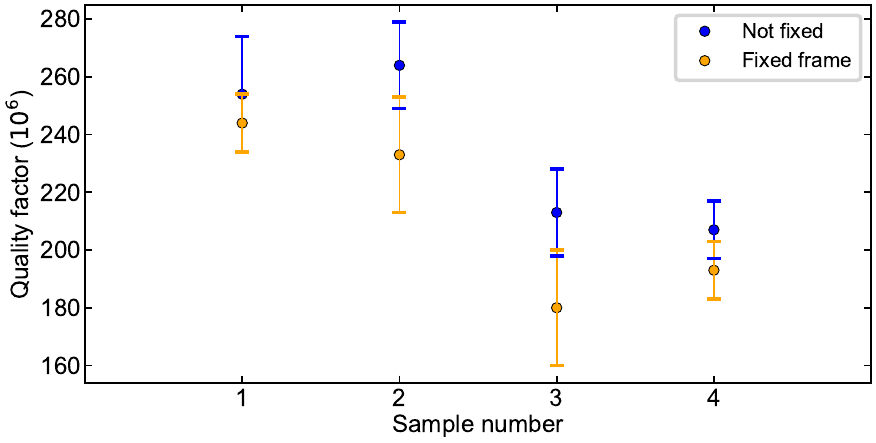}
    \caption{\textbf{Effect of chip mounting on the quality factor.} Blue dots show the measured $Q$ when the chip is freely mounted on the sample holder and orange dots correspond to the $Q$ when fixing the chip tightly using metallic clamps pressing on the frame.}
    \label{fig_SI:clamping effect}
\end{figure}

The quality factors of nanomechanical resonators generally experience degradation (clamping losses) when the chip on which they are fabricated is fixed by the application of a force \cite{schmid2011damping,wilson2009cavity}. To investigate this degradation, we first leave the chip unclamped on the characterization holder, then clamp the chip tightly using metallic clamps pressing on the edges of the chip and fixed with screws and measure the quality factor in both cases. We characterize polygons with $l_0 = \SI{150}{\micro\meter}$, $r_l=0.6$ and $w_0 = \SI{300}{\nano\meter}$ and compare the measured $Q$. We measure 4 different devices and observe $1-10\%$ degradation in the $Q$ for the fixed frame case. Error bars in \fref{fig_SI:clamping effect}{} show the standard deviation of 4 measurements for each sample. These measurements show that the degradation due to chip clamping conditions are substantially suppressed, as previous studies with unpatterned nanobeams and membranes show substantially larger degradation \cite{schmid2011damping,wilson2009cavity}

\section{Strain engineered polygon resonators}\label{appendix: stress_engineering}
\begin{figure}[t]
	\includegraphics[width=\linewidth]{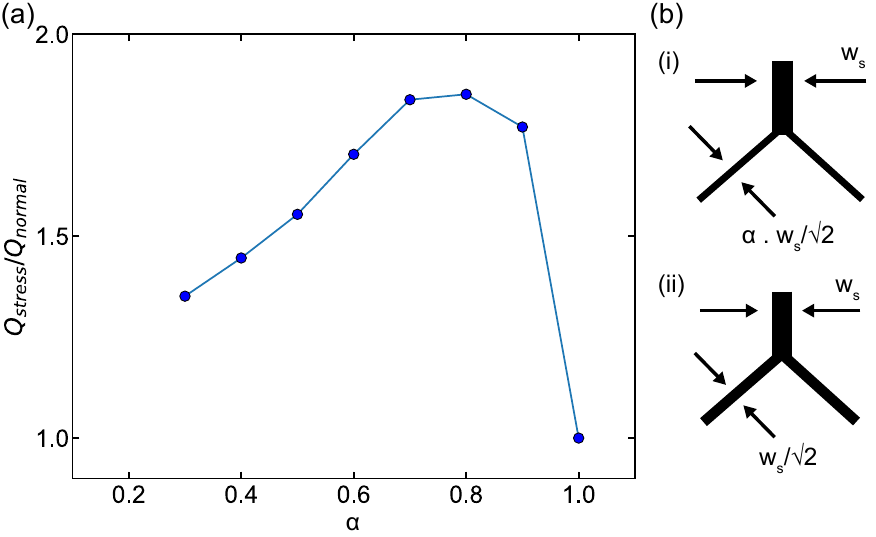}
	\caption{\textbf{Strain engineered polygon resonators.} (a) Simulated $Q$ enhancement of polygon resonators via strain engineering ($Q_\mathrm{stress}$) compared to the stress preserving design ($Q_\mathrm{normal}$). (b) Schematic of the polygon width tapering for strain engineering (i) and stress preserved design (ii) for a square-shaped polygon.}
	\label{fig:Stress_Engineering}
\end{figure}

Dissipation dilution suggests that one method to improve the $Q$ of a soft clamped mode is to enhance the local stress in the region where the mode is localized \cite{ghadimi2018elastic}. This method---strain engineering---has shown $Q$ enhancement in nanobeam resonators beyond the soft clamping limit by width tapering the nanobeam to increase the stress in the localized mode region. We investigate strain engineering in polygon resonators in simulations. By keeping the support width ($w_s$) constant and narrowing the polygon side width ($w_0$) by a tapering constant $\alpha<1$, we can enhance the local stress in the sides of the polygon. For a suspended thin film with a fixed deposition stress, the total tension force at the junction is constant and by narrowing the sides of the polygon stress is enhanced (\eqref{eq:sigma_SI}). We simulate the $Q$ of the strain engineered design shown in \fref{fig:Stress_Engineering}{(b)} panel (i) ($Q_\mathrm{stress}$) and compare it to the stress preserving design ($Q_\mathrm{normal}$) that is studied in this work (\fref{fig:Stress_Engineering}{(b)} panel (ii)). We simulate a square with $l_0 = \SI{250}{\micro\meter}$, $r=0.6$, $w_0 = \SI{200}{\nano\meter}$ and $\SI{20}{\nano\meter}$ thickness. The simulation results in \fref{fig:Stress_Engineering}{(a)} show possible $Q$ enhancements up to $1.8$ times higher than the stress preserved design. 

\section{Fabrication process flow}\label{appendix: fab}

We fabricate all the samples using $\SI{20}{\nano\meter}$ thick LPCVD \SiN\ on Si wafers. Structures are defined using electron beam (ebeam) lithography using FOX16 resist. Patterns are transferred into the \SiN\ using $\mathrm{SF_6}$ based dry etching. A biased mask is written using the same ebeam lithography process and used for deep reactive ion etching (DRIE) of the silicon substrate to create a recess for the final suspension etch. Samples are then diced to chips and cleaned using Piranha and BHF solutions prior to the final etch. \SiN\ structures are released using KOH wet etching and dried using a critical point dryer (CPD). Full details of the nanofabrication process is available on NanoFab fabrication archive \cite{Groth_Bereyhi_2020}.
\begin{figure}[h]
    \centering
    \includegraphics[width=\linewidth]{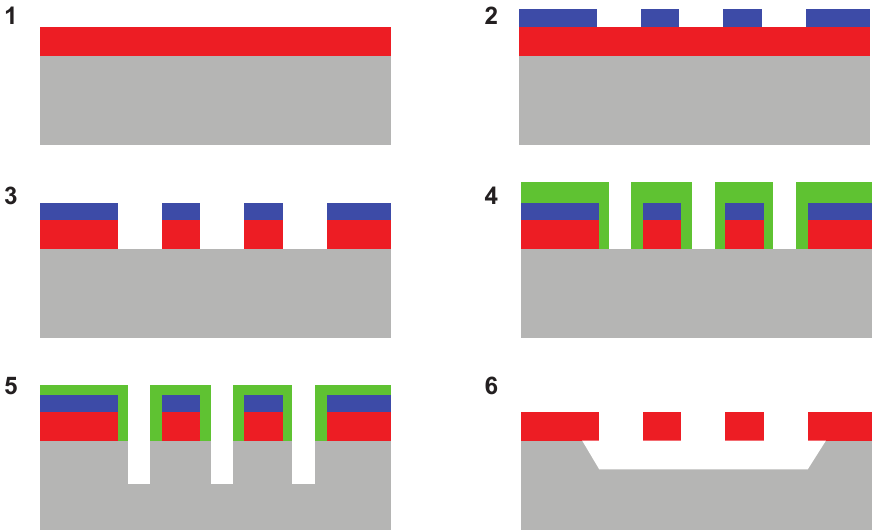}
    \caption{\textbf{Fabrication process flow.} \textbf{1} LPCVD thin film deposition of \SiN  on Si (red on gray). \textbf{2} Ebeam lithography on \SiN wafer using FOX16 (blue). \textbf{3} Dry etching of \SiN. \textbf{4} Recess ebeam (green corresponds to the second layer of FOX16). \textbf{5} DRIE etching of Si substrate. \textbf{6} KOH undercut and CPD.}
    \label{fig_SI:PF}
\end{figure}

\section{Frequency stability of fabricated mechanical resonators}\label{appendix: freq_stability}

For many sensing applications, the frequency stability is a relevant figure of merit. Therefore, we measured the frequency stability of these mechanical resonators by characterizing their Allan deviation. The Allan deviation can be derived in terms of the frequency noise spectrum $S_{\Omega_m}$~\cite{gavartin2013stabilization},
\begin{equation}
\begin{split}
	&\sigma^2(\tau) = \frac{1}{2\Omega_m^2}\frac{1}{N-1}\sum_{k=2}^{N}(\overline{f}_k - \overline{f}_{k-1})^2 = \\  &\frac{2}{\pi}(\frac{2}{\Omega_m\tau})^2\int_{-\infty}^{\infty}d\omega S_{\Omega_m}[\omega]\sin^4(\omega\tau/2)/\omega^2 = \\ &\frac{2}{\pi}(\frac{2}{\Omega_m\tau})^2\int_{-\infty}^{\infty}d\omega S_{\phi}[\omega]\sin^4(\omega\tau/2)
\end{split}
\end{equation}
where $\overline{f_k}$ is the averaged frequency of the oscillator over the $k$th measurement fragment of duration $\tau$, and the relation $S_{\Omega_m}[\omega] = \omega^2S_{\phi}[\omega]$ is used.
The thermomechanical phase noise of the oscillator in the strong drive limit is given by
\begin{equation}
	S_\phi[\omega] = \frac{1}{\langle X_\mathrm{osc}^2\rangle}\frac{ k_BT}{m\Omega_m^2}\frac{\Gamma/2}{\omega^2+(\Gamma/2)^2} = \frac{\langle X_\mathrm{th}^2\rangle}{\langle X_\mathrm{osc}^2\rangle}\frac{\Gamma/2}{\omega^2+(\Gamma/2)^2}.
\end{equation}
therefore, the Allan deviation can be simplified to
\begin{equation}
	\sigma^2(\tau) = \frac{\langle X_\mathrm{th}^2\rangle}{\langle X_\mathrm{osc}^2\rangle} \frac{1}{Q_m\Omega_m\tau}\frac{(\tau\Gamma_m)^2}{1+(\tau\Gamma_m)^2}
\end{equation}
The other noise sources are the the detection noise and the long time linear drift. The detection noise $S_\phi^d(\omega) = \frac{\Delta f S_{\mathrm{Noise}}}{\langle X_\mathrm{osc}^2\rangle}\frac{\Delta f/2}{\omega^2+(\Delta f/2)^2}$ is assumed to arise from a flat background noise $S_{\mathrm{Noise}}$ in the measurement record, limited only by the detection bandwidth $\Delta f$. When $\tau$ is much longer than the detection time ($\sim1/\Delta f$), the Allen deviation caused by the detection noise is $\sigma_d(\tau) = \sqrt{\frac{\Delta f S_{\mathrm{Noise}} }{\langle X_\mathrm{osc}^2\rangle}\frac{2\pi\Delta f}{\Omega_m^2\tau}}$. For the long time linear drift $\delta \Omega_m/\Omega_m = Dt$, the Allan deviation is $\sigma_l(\tau) = \frac{D\tau}{\sqrt{2}}$.  \\

\begin{figure}[!h]
\centering
\includegraphics[width=\linewidth]{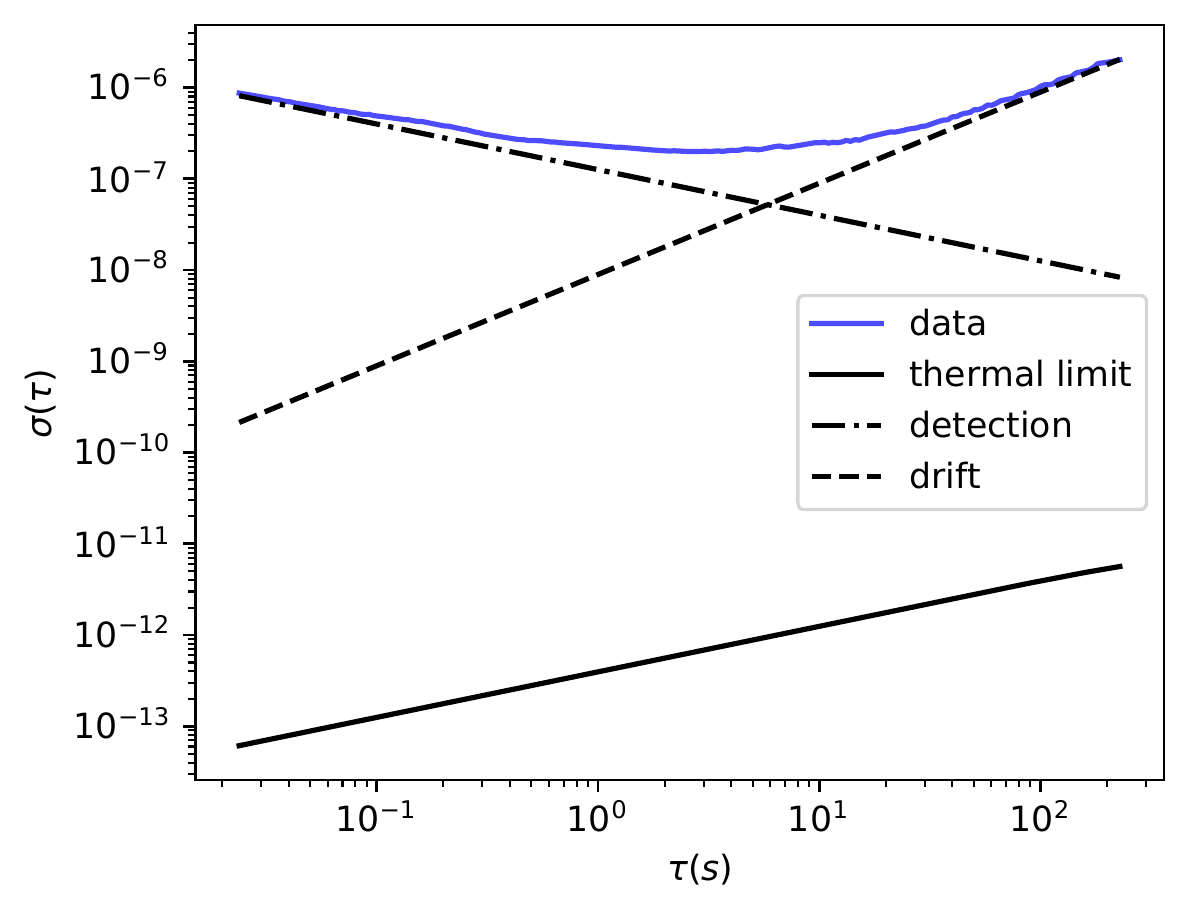}
\caption{\textbf{Frequency stability of perimeter modes.} The measurement result of Allan deviation of a $\SI{1}{\milli\meter}$ square-shaped sample with $w_0 = \SI{200}{\nano\meter}$ and $r_l = 0.2$ is plotted in blue dots. The dash-dotted line shows the calculated detection limit based on the measurement SNR, and the dashed line is the fit to the linear frequency drift in the long time limit. The fit yields a drift rate $D \sim \SI{e-8}{s^{-1}}$, which varies among different samples. The theoretical thermal mechanical limit is plotted in black line, which is many orders of magnitude below the measured Allan deviation, due to the exceptionally high quality factor. The scaling is reversed from the usual $\tau^{-1/2}$ to $\tau^{1/2}$ because we are in the limit where the mechanics life time is longer than the measurement time $\tau < 1/\Gamma_m$. }
\label{fig:1}
\end{figure}
To investigate the frequency stability of our resonators, we use a phase-locked loop (PLL) with a bandwidth of \SI{1}{kHz} to track the mechanical frequency fluctuations in our optical interferometer. The measurement result is shown in Fig.~\ref{fig:1}. The result indicates that the mechanical frequency stability is limited by the linear drift (rate $D \sim \SI{e-8}{s^{-1}}$) at long $\tau$ limit, which masks the extremely low thermomechanical noise-induced frequency fluctuations (Fig.~\ref{fig:1}), many orders of magnitude below. The rate of this linear drift varies between different samples and is consistent with the long term drift in the resonance frequency observed over 24 hours. As the frequency drift is always positive (i.e. increases the resonance frequency), it could be caused by mass reduction due to evaporation of a volatile contaminant on the resonator inside the vacuum chamber.

\section{Simulations of the phononic dimers}\label{sec:SI_dimers_simulations}

\begin{figure}[t]
        \centering
    \includegraphics[width=\linewidth]{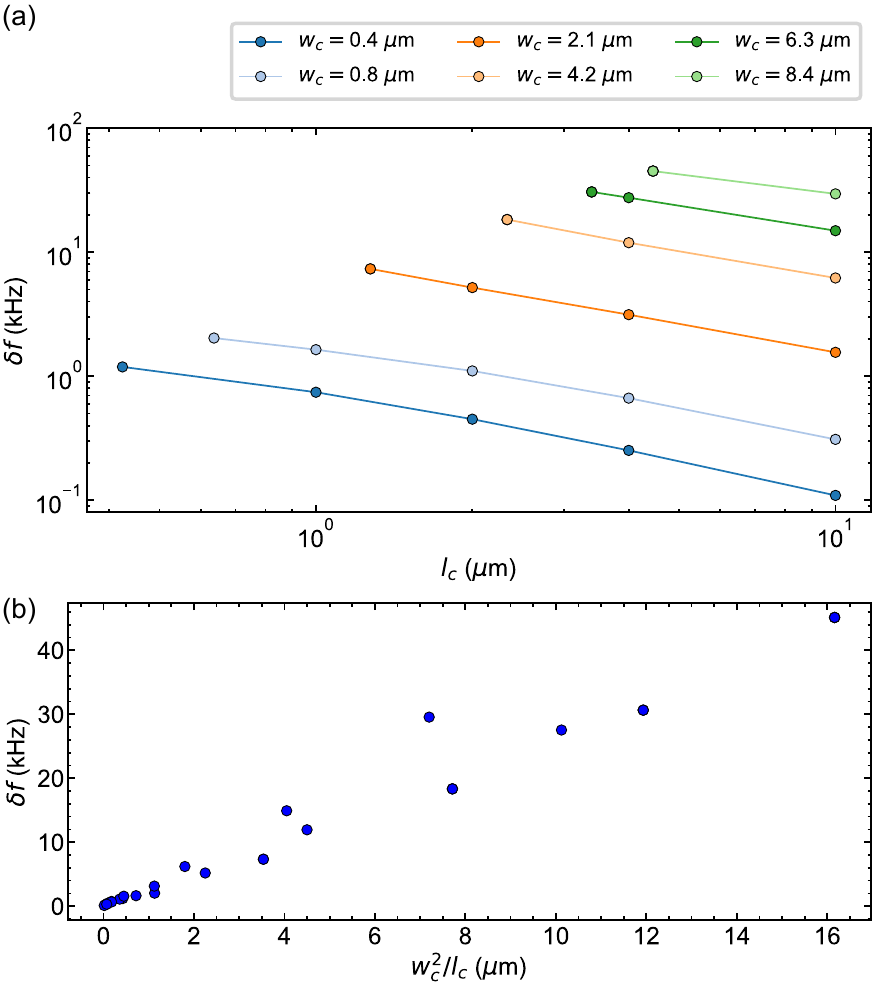}
    \caption{\textbf{Frequency splitting of dimers.} (a) FEM simulation of mode splitting for varying coupler length ($l_c$) and different coupler widths ($w_c$). (a) Compilation of all the simulated values in panel (a) plotted versus $w_c^2/l_c$}
    \label{fig_SI:Dimer_wc2}
\end{figure}

We simulate the mode splitting of the dimers for squares of $\SI{150}{\micro\meter}$ side length, $\SI{300}{\nano\meter}$ side width and $r=0.2$. We sweep the length of the coupler for different coupler widths as shown in \fref{fig_SI:Dimer_wc2}{(a)}. We find that the mode splitting depends approximately linearly on  $w_c^2/l_c$, where $w_c$ is the width of the coupling segment and $l_c$ is the length of the coupling segment (\fref{fig_SI:Dimer_wc2}{(b)}). Since the coupler width dependence of the mode splitting is stronger than the length dependence we choose the shortest and widest possible couplers to overcome the disorder in the dimers and chains. The minimum coupler length for a given width is determined by the requirement that the coupler segment must be attached on all four corners. Therefore, for wider couplers compared to the beam width, we adapt the coupler length such as to avoid buckling and deformation of the coupling segment after device suspension.

\section{Full mode spectra of the phononic arrays}\label{sec:SI_arrays_full_spectra}

\begin{figure}[t]
    \centering
    \includegraphics[width=\linewidth]{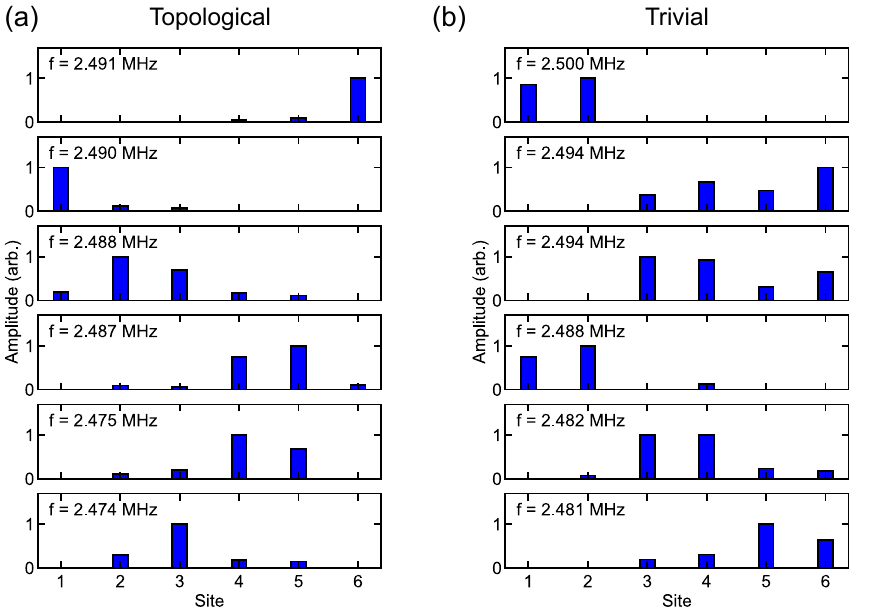}
    \caption{\textbf{Mode profiles of the collective perimeter modes.} RMS values of the Brownian motion of each site measured for all the collective perimeter modes of both topological (a) and trivial (b) chains of six square-shaped resonators.}
    \label{fig_SI:chain_mode_profiles}
\end{figure}

A chain of $n$ polygon resonators has many flexural modes and we only consider the perimeter modes of the structure. We identify $n$ perimeter modes of the chain by excluding modes with non-zero amplitude on the supporting tethers. Using the same approach as explained in the main text, we find the mode profiles of the collective perimeter modes in the spectrum of both topological and trivial chains. These modes have frequencies in the proximity of the perimeter mode of a single site and have no amplitude on the tethers (verifying that they are localized on the perimeters). These mode profiles are shown in \fref{fig_SI:chain_mode_profiles}{}. One can observe that the edge modes are only present in the topological case. Nevertheless, the profile of the other modes as well as the order of their appearance in the spectrum is dominated by the disorder in the fabricated devices.

\section{Disorder in phononic dimers and arrays}
\begin{figure*}[t]
    \centering
    \includegraphics[width=\linewidth]{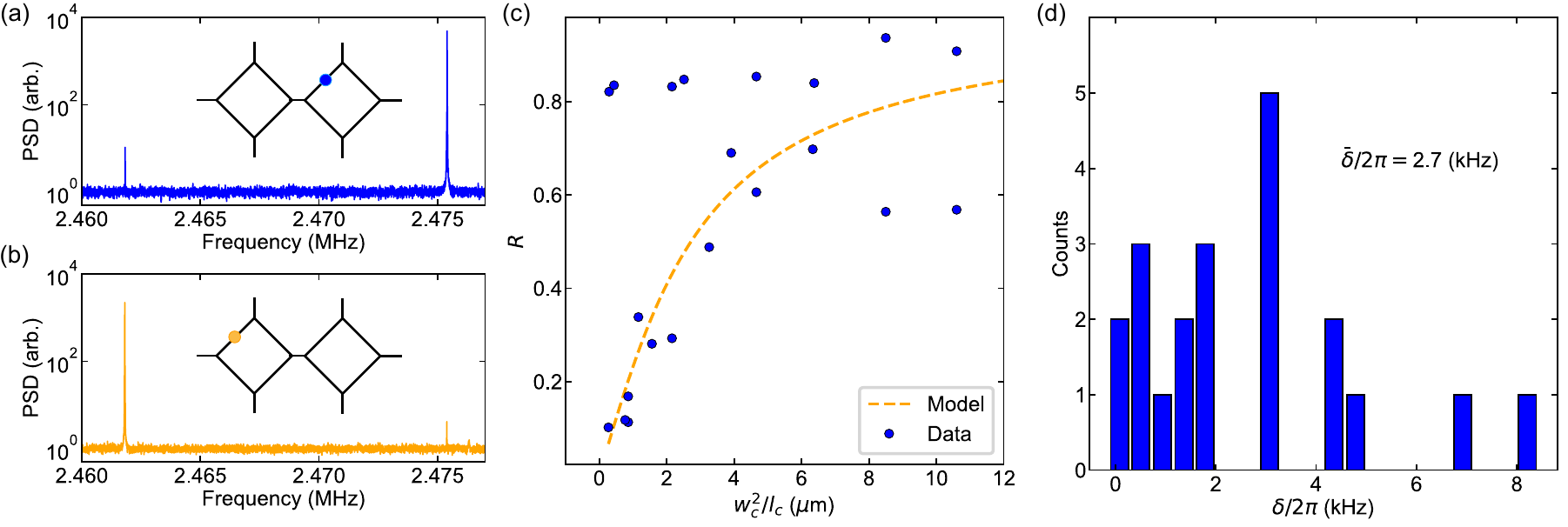}
    \caption{\textbf{Disorder in phononic dimers and arrays.} (a), (b) Spectra of the interferometric measurement when the laser beam is focused on sides of the two resonators in a phononic dimer. The dimer's dimensions are similar to the ones shown in \fref{fig:Dimer_Chain}{} with $w_c = \SI{2.1}{\micro\meter}$ and $l_c = \SI{11.1}{\micro\meter}$ corresponding to nominal frequency splitting of $\SI{1.1}{\kilo\hertz}$. (c) Asymmetry ratio (\eqref{eq_SI:Asym_ratio}) obtained from the experimental spectra for devices with different coupling parameters, used for the measurement in \fref{fig:Dimer_Chain}{(e)}. The dashed line is the model \eqref{eq_SI:Asym_ratio} with $\delta$ set as its average value. (d) Histogram of the disorder parameter $\delta$ extracted from the data shown in panel c using \eqref{eq_SI:delta_sym_ratio}}.
    \label{fig_SI:dimer_disorder}
\end{figure*}
Fabrication imperfections cause dimer and array devices to deviate from an ideal model of coupled identical resonators. These deviations manifest as asymmetry in the hybridized modes of dimers or loss of topological protection in arrays. In reality the disorder can be the result of various processes (e.g. inhomogeneous stress distribution or imperfections in the lithography step). However, one can reduce all the imperfections to a difference between the bare frequencies of the resonators and characterize the disorder using a simple model of coupled harmonic oscillators. In particular, for an 'imperfect' dimer, the Hamiltonian can be written in the frame rotating with the average frequency of the two resonators as
\begin{equation}
    H = \begin{pmatrix} \delta & J \\ J & -\delta \end{pmatrix},
\end{equation}
where $J$ is the coupling rate and $\delta$ is the absolute value of half of the difference between the bare frequencies. The eigenfrequencies of this Hamiltonian are given by $\omega_{\pm} = \pm\sqrt{\delta^2 + J^2}$ corresponding to eigenvectors 
\begin{align}
    V_+ &= \begin{pmatrix}  J \\ \sqrt{J^2+\delta^2}-\delta \end{pmatrix}\label{eq_SI:V_plus},\\
    V_- &= \begin{pmatrix}  \sqrt{J^2+\delta^2}-\delta \\ -J \end{pmatrix}\label{eq_SI:V_minus},
\end{align}
where $+$ and $-$ correspond to in-phase and out-of-phase modes respectively. Here we observe that the disorder parameter $\delta$ not only contributes to the frequency splitting, but also results in an asymmetry between the amplitudes of perimeter modes of each resonator in the hybridized modes. In our experiment, this asymmetry manifests as a difference between the detected SNR of the hybridized modes when the interferometer is focused on sides of the two resonators (\fref{fig_SI:dimer_disorder}{(a, b)}). According to \eqref{eq_SI:V_plus} and \eqref{eq_SI:V_minus}, the ratio of the smaller amplitude to the larger one is given by
\begin{equation}\label{eq_SI:Asym_ratio}
    R = \sqrt{1 + \frac{\delta^2}{J^2}} - \frac{\delta}{J}.
\end{equation}
We measure this ratio by taking the ratio of RMS thermal fluctuations. We observe that as expected from \eqref{eq_SI:Asym_ratio} for devices with higher coupling parameters, the ratio is closer to one (\fref{fig_SI:dimer_disorder}{(c)}). One can also reverse \eqref{eq_SI:Asym_ratio} and find the disorder parameter as a function of the asymmetry ratio
\begin{equation}
    \frac{\delta}{J} = \frac{1-R^2}{2R}\label{eq_SI:delta_sym_ratio}.
\end{equation}
Using the linear relation found in \fref{fig_SI:Dimer_wc2}{(b)}, we can compute $J$ from the geometric design parameters and translate \eqref{eq_SI:delta_sym_ratio} to actual frequency and find the distribution of the disorder parameter for these devices. The histogram for $\delta$ is shown \fref{fig_SI:dimer_disorder}{(d)}) and has an average around $\bar\delta/2\pi = \SI{2.7}{\kilo\hertz}$ (equivalent to frequency splitting of \SI{5.4}{\kilo\hertz}). By plugging this average value into \eqref{eq_SI:Asym_ratio} we can explain the average trend of the asymmetry ratio as a function of the coupling parameter (\fref{fig_SI:dimer_disorder}{(c)}). These results explain the deviations from the FEM simulation in \fref{fig:Dimer_Chain}{(e)} for frequency splittings below \SI{10}{\kilo\hertz} as well as the localization in the edge modes of our phononic array (\fref{fig:Dimer_Chain}{(h)}).

%

\end{document}